\documentclass[aip,reprint,floatfix,twocolumns,cha]{revtex4-2}
\usepackage{amsfonts,amssymb,amsmath,array}
\usepackage{comment}
\usepackage{cancel}
\usepackage{bm}
\usepackage[normalem]{ulem}

\usepackage{hyperref}
\hypersetup{colorlinks=true, linkcolor=black, citecolor=black, urlcolor=black}

\usepackage{lipsum}
\usepackage[final]{graphicx}
\usepackage{graphicx}
\usepackage{epstopdf}
\usepackage{mathtools}
\graphicspath{{./Figures/}} 

\usepackage[usenames]{color} 
\usepackage{xcolor}
\usepackage{soul} 

\begin{document}

\title{Curvature-driven wall accumulation in chiral active particles}


\author{Alessandro Petrini}
\affiliation{Sapienza University of Rome, P.le A. Moro 2, Rome, Italy}

\author{Raphaël Maire}
\affiliation{Department of Condensed Matter, University of Barcelona, 08028 Barcelona, Spain}

\author{Umberto Marini Bettolo Marconi}
\affiliation{School of Sciences and Technology, University of Camerino, Via Madonna delle Carceri, Italy}

\author{Lorenzo Caprini}
\affiliation{Sapienza University of Rome, P.le A. Moro 2, Rome, Italy}
\email{lorenzo.caprini@uniroma1.it}

\newcommand{\bea}{\begin{eqnarray}}   
\newcommand{\eea}{\end{eqnarray}}

\newcommand{\beq}{\begin{equation}} 
\newcommand{\eeq}{\end{equation}}

\newcommand{\er}{{\hat{{\bf r}}}}
\newcommand{\etheta}{{\hat{{\boldsymbol{\theta}}}}}
\newcommand{\en}{{\hat{{\bf n}}}}
\newcommand{\et}{{\hat{{\bf t}}}}
\newcommand{\ex}{{\hat{{\bf x}}}}
\newcommand{\ey}{{\hat{{\bf y}}}}
\newcommand{\ez}{{\hat{{\bf z}}}}
\newcommand{\norm}[1]{\left\| #1 \right\|}
\newcommand{\kB}{k_{\rm B}}
\newcommand{\eps}{\varepsilon}

\newcommand{\DD}{\mathbf{D}}
\newcommand{\FF}{\mathbf{F}}
\newcommand{\TT}{\mathbf{T}}
\newcommand{\rr}{\mathbf{r}}
\newcommand{\uu}{\mathbf{u}}
\newcommand{\vv}{\mathbf{v}}
\newcommand{\ww}{\mathbf{w}}
\newcommand{\xx}{\mathbf{x}}

\newcommand{\pr}{^{\,\prime}}
\newcommand{\ppr}{^{\,\prime\prime}}

\newcommand{\newpar}{\mathbin{\!/\mkern-5mu/\!}}

\newcommand{\bnabvn}{{\bfnabla_{{\bf v}_i}}}
\newcommand{\bnabrn}{{\bfnabla_{{\bf r}_i}}}
\newcommand{\bnabri}{{\bfnabla_{{\bf r}_i}}}
\newcommand{\bnabr}{{\bfnabla_{{\bf r}}}}
\newcommand{\bnabv}{{\bfnabla_{{\bf v}}}}
\newcommand{\bnabeta}{{\bfnabla_{{}}}}
\newcommand{\bfnabla}{\mbox{\boldmath $\nabla$}}
\newcommand{\bnabvin}{{\bfnabla_{{\bf v}_i}}}

\newcommand{\unittheta}{\boldsymbol{\hat{\theta}}}
\newcommand{\bfxi}{\boldsymbol{\xi}}
 \newcommand{\bepsilon}{\boldsymbol{\epsilon}}
\newcommand{\kk}{\boldsymbol{\kappa}}
\newcommand{\eeta}{\boldsymbol{\eta}}
\newcommand{\xxi}{\boldsymbol{\xi}}
\newcommand{\cchi}{\boldsymbol{\chi}}
\newcommand{\bomega}{\boldsymbol{\Omega}}
\newcommand{\smallomega}{\boldsymbol{\omega}}
\date{\today}

\newcommand{\comm}[1]{\textcolor{red}{ #1}}
\newcommand{\R}[1]{\textcolor{red}{ #1}}
\newcommand{\edit}[1]{\textcolor{blue}{ #1}}

\begin{abstract}


We study a dilute system of non-motile chiral active particles confined in geometries ranging from straight channels to circular enclosures.
Activity is introduced through chiral particle–wall interactions, modeled as tangential wall forces that generate the edge currents characteristic of chiral active matter. Remarkably, although the particles lack self-propulsion, these boundary currents induce density inhomogeneities. We show that boundary curvature drives a wall accumulation phenomenon: particles remain uniformly distributed in straight channels but accumulate near the boundaries of circular confinements. Numerical simulations and a hydrodynamic theory for the density and momentum fields consistently capture this curvature-induced wall-accumulation. These results identify boundary curvature as a fundamental control parameter for chiral edge transport and confinement-induced organization, with potential experimental relevance to spinning colloids and granular spinners.

\end{abstract}

\date{\today}

\maketitle


\section{Introduction}\label{ch:Intro}


Active matter encompasses microscopic and macroscopic systems whose constituents continuously convert energy from their surroundings into persistent motion or collective effects~\cite{marchetti2013hydrodynamics, elgeti2015physics, bechinger2016active}. A broad class of active systems exhibits chirality~\cite{lowen2016chirality}, arising either from rotationally asymmetric particle shapes or propulsion mechanisms~\cite{liebchen2022chiral}. As a consequence, chiral active particles often follow circular or helical trajectories and display motion and transport properties that differ fundamentally from those of their non-chiral counterparts~\cite{fruchart2023odd}. Experimental realizations include L-shaped~\cite{kummel2013circular} or rotating colloids~\cite{mecke2024chiral}, bacteria~\cite{petroff2015fast}, spermatozoa~\cite{woolley2003motility}, cells~\cite{xu2007polarity} algae~\cite{drescher2009dancing,huang2021circular}, and other microorganisms~\cite{tan2022odd}, as well as droplets~\cite{carenza2019rotation} and macroscopic active granular systems composed of chiral vibrobots~\cite{caprini2025spontaneous,scholz2021surfactants,kiechl2026free}, light-driven circular walkers~\cite{siebers2023exploiting}, or spinners~\cite{scholz2018rotating, lopez2022chirality, carrillo2025depinning}. 

Chirality can affect the dynamics of active systems in several distinct ways. On the one hand, it is often associated with self-propulsion mechanisms combining a persistent propulsion speed with a finite angular velocity, resulting in curved particle trajectories~\cite{van2008dynamics}. On the other hand, chirality can generate effective interactions that are intrinsically non-conservative~\cite{caprini2026modeling,huang2025anomalous,maire2026kinetic,guo2026tuning}. In particular, it may induce odd interactions~\cite{caprini2025bubble}, namely forces acting transversely to the line connecting two interacting particles or to the local wall normal in the case of particle-wall interactions. Such forces, previously introduced as curl forces~\cite{berry2012classical}, cannot be derived from an interaction potential and therefore have no equilibrium counterpart. Effective odd interactions naturally emerge from coarse-grained descriptions of microscopic mechanisms, including tangential friction in granular spinner systems~\cite{caprini2026modeling, digregorio2026phase} and hydrodynamic interactions between magnetically driven rotating colloids~\cite{massana2021arrested, caprini2025odd}.

Most theoretical studies of chiral active matter have focused on motile (self-propelled) particles. Even in the dilute limit, motile chiral particles exhibit a variety of nonequilibrium phenomena with no passive counterpart. At the single-particle level, chirality modifies transport properties, leading to oscillatory mean-square displacements~\cite{van2008dynamics,barman2025confinement,caprini2025active}, reduced long-time diffusivities~\cite{sevilla2016diffusion, olsen2024optimal}, and odd diffusion~\cite{hargus2021odd, vega2022diffusive, kalz2022collisions, mecke2025obstacle, abdoli2026dynamical, kalz2024oscillatory, hargus2025passive,marini2026emergent} characterized by antisymmetric off-diagonal components of the diffusion tensor. Under confinement, chiral active particles satisfy an equation of state as verified experimentally through chiral vibrobots~\cite{caprini2025active}, relating mechanical pressure and density through an effective temperature. Furthermore, chirality generates persistent currents flowing tangentially to harmonic potentials~\cite{caprini2023chiral} or confining boundaries~\cite{caprini2019active}, commonly referred to as edge currents~\cite{van2016spatiotemporal,adorjani2024phase,li2024robust,wang2026edge,alsallom2026origin,caprini2025active}, which modify the transport properties in channel geometries~\cite{iyaniwura2026splitting,khatri2026diffusion}.
A robust consequence of self-propulsion is the accumulation of active particles near obstacles and confining walls~\cite{elgeti2013wall, caprini2018active}. This phenomenon occurs in both chiral~\cite{caprini2019active} and non-chiral systems~\cite{caprini2020activity, vsindelka2025confined} and is generally attributed to the persistence of active motion~\cite{maggi2015multidimensional}: particles reaching a boundary continue to propel against it until their orientation decorrelates~\cite{lee2013active,duzgun2018active,das2018confined,wagner2022steady}. As a result, wall accumulation is typically regarded as a direct consequence of motility.

In this work, we show that wall accumulation can emerge even in non-motile active particles as a direct consequence of edge currents generated by chiral, non-conservative particle-wall interactions. In contrast to the well-known motility-induced accumulation mechanism, the effect reported here is controlled by the curvature of the confining boundary. Particles remain uniformly distributed in geometries bounded by flat walls, where the curvature vanishes, whereas they accumulate near the boundaries of circular confinements. The physical origin of this curvature-driven accumulation lies in the edge currents themselves: along curved boundaries, particles follow curved trajectories and experience an effective centrifugal force that is absent in flat geometries. This force drives the accumulation, leading to inhomogeneous density profiles. We characterize this phenomenon through numerical simulations and develop a hydrodynamic theory for the density and momentum fields that predicts the observed behavior.

The remainder of the paper is organized as follows. In Sec.~\ref{ch:Model}, we introduce the model. In Sec.~\ref{ch:WallAccumulation}, we present the numerical results and characterize the curvature-induced accumulation mechanism, while Sec.~\ref{ch:EdgeCurrents} characterizes the associated edge currents and in Sec.~\ref{ch:Theory}, we derive the corresponding hydrodynamic theory. Finally, Sec.~\ref{ch:conclusions} summarizes our conclusions and discusses future perspectives.

\section{The model for chiral walls}\label{ch:Model}

\begin{figure*}[t!]
\includegraphics[width=0.98\textwidth, trim=0 0 0 0, clip=true]{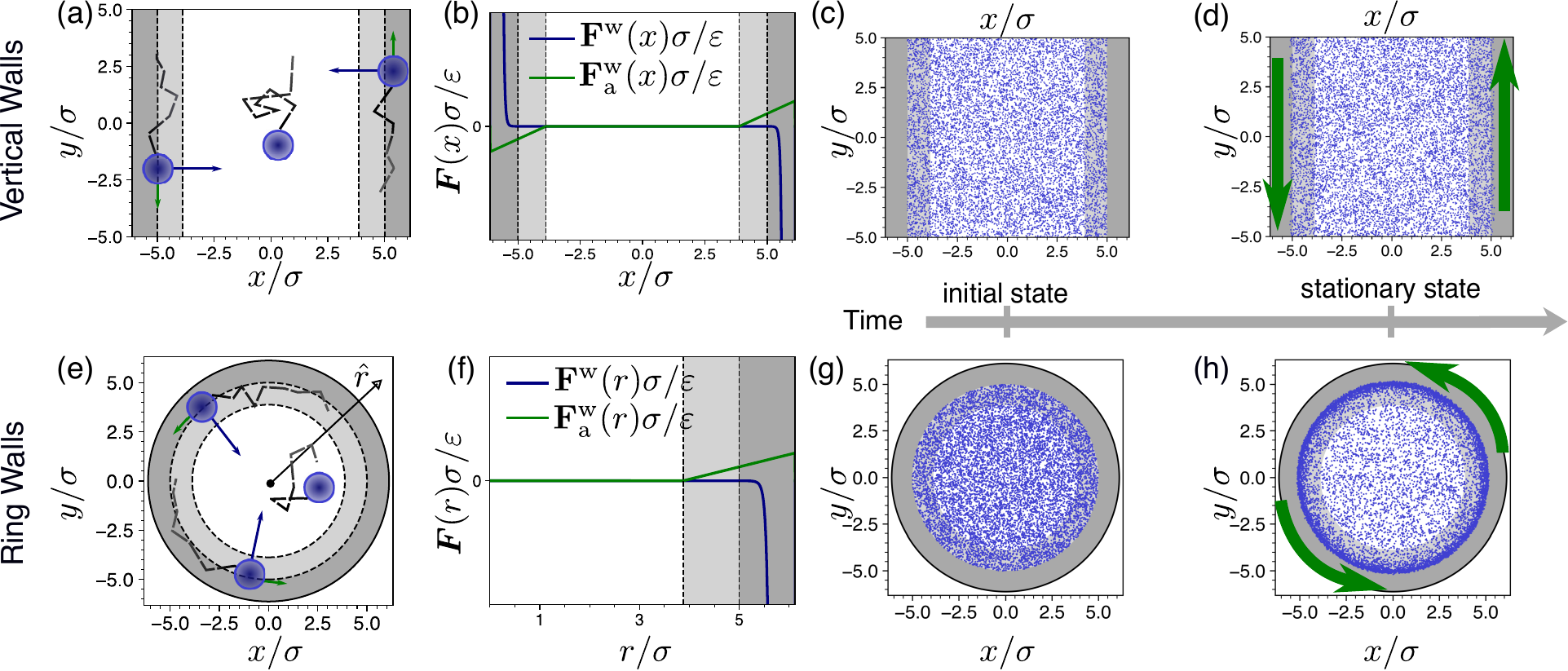}
\caption{{\textbf{Density accumulation near wall for Brownian particles confined with chiral interactions}}. The figure shows the effect of density accumulation at the wall due to
curvature, comparing the vertical wall ((a)-(d)\,) geometry to the ring wall ((e)-(g)\,) geometry . Figures (a)-(e) display typical trajectories for three particles inside the simulation box: inside the bulk region (white area), particles move erratically, performing Brownian motion, while inside the grey-shaded area they feel a repulsive force term and a transverse one, which makes the particles move tangentially along the wall, stabilizing at a distance from it. Inside the light-grey-shaded area, particles can feel only the transverse term, while inside the dark-grey region they also feel the repulsive force, which acts along the normal to the wall. Figures (b)-(f) represent a schematization of the intensity of the wall force, $\FF^{\rm{w}}$, and the active chiral force, $\FF_{\rm{a}}^{\rm{w}}$, inside the simulation box, varying the normal component: inside the bulk (white) region the wall forces are zero, in the light-grey shaded region the chiral force becomes non-zero, while the repulsive force is present only in the dark-grey region. Signs due to the normal direction are included in the plots. We recall that in the case of vertical wall geometry the normal component is defined as the distance between the particle and the wall along the $x$-direction, while in the case of the ring wall along the radial direction. 
Snapshots (c)-(g) represent a population of  $N=10^{4}$ independent and non-interacting particles for the two wall geometries in the homogeneous state, which is used as an initial condition for the dynamics. Snapshots (d) and (h) display the stationary states, respectively for the systems (c) and (g): in the ring wall geometry one can notice density accumulation near the wall, while in the vertical wall geometry no accumulation is noticeable. Both systems display currents as depicted by the black arrows. Snapshots (c)-(d) and (g)-(h) refer to simulations realized using $\tau\gamma/m = 100$, $\alpha_{0}\sigma^{2}/\varepsilon = 500$, and $T/\varepsilon = 1.0$ in units of $k_{\mathrm{B}}$ and time units defined as $\tau = \sigma \sqrt{m/\eps}$.
}
\label{fig:Fig0}
\end{figure*}

Here, we consider the dynamics of non-interacting chiral active particles confined within different geometries. 
Activity is included in the interactions between each particle and the wall, which not only repels the particles, acting normal to the wall profile, but also induces an additional force tangential to its profile. This effective force naturally appears when a rotating object encounters a wall: it is caused by the tangential friction when a granular spinner touches a wall, while it is effectively due to hydrodynamic interactions when a spinning colloid is close to a wall.

The particle dynamics in two dimensions is modelled through an evolution equation for the particle position, $\mathbf{r}$, and velocity, $\vv=\dot{\rr}$:
\beq
m\dfrac{d\vv}{dt} = -\gamma\vv + \sqrt{2\gamma T}\,\xxi + \FF^{\rm{w}} + \mathbf{F}^{\rm{w}}_{\rm{a}}\,\rm{,}
\label{eq:langevin_eq}
\eeq
where $\xxi$ is a Gaussian white noise with unit variance.
Here, $m$ is the particle mass, while $\gamma$ and $T$ represent the drag coefficient and the temperature of the bath (in units of $\kB$), respectively. 

The particles are confined to move in a closed domain surrounded by a wall with a characteristic profile represented by the curve $\ww(s)$, where $s$ is the arc-length parameter. 
In the following, we focus on flat (vertical-wall) and ring (circular confinement) geometries. Without loss of generality, we can describe both systems using the tubular-coordinate representation for the particle position, $\rr$:
\beq
\rr(s,n) = \ww(s) + n\,\en (s)\,\rm{.}
\label{eq:tc_embedding}
\eeq
In this coordinate system, the particle position is decomposed into a component tangential to the wall, following the profile $\ww(s)$, and a component $n$ indicating the distance from the wall along the wall-normal direction $\en(s)$.
For instance, in the case of a vertical wall aligned along the y-direction, the parameter $n$ is the difference between the $x$-coordinate of the wall and that of the particle, whereas in the case of ring walls, {\color{blue} $n$} is defined as the radial distance between the wall and the particle.

In dynamics~\eqref{eq:langevin_eq}, the terms $\FF^{\rm{w}}$ and $\mathbf{F}^{\rm{w}}_{a}$ represent the forces generated by the wall and are distinguished into two contributions. 
The former, $\FF^{\rm{w}}$, accounts for volume exclusion effects through a repulsion term acting along the wall-normal direction, $\en(s)$, pointing from the wall to the inner region.
This force is conservative, as in passive systems, and depends only on the distance $n$ between the particle and the wall, and can be expressed as
\begin{equation}
\mathbf{F}^{\rm{w}}(n) = -\partial_{n} U^{\rm{w}}(n)\,\Theta({n}_{c}-n) \hat{\mathbf{n}}
\label{eq:wall_force_n}     
\end{equation}
where $\Theta$ is the Heaviside step function and $n_{c}$ is the cut-off distance of the force. The term $U^{\rm{w}}$ is the Weeks-Chandler-Andersen potential with the form 
\begin{equation}
U^{\rm{w}}(n)=4\eps\left[\left(\frac{\sigma}{n}\right)^{12}-\left(\frac{\sigma}{n}\right)^{6}\right]+\eps \,\rm{.}
\label{eq:WCA_pot}
\end{equation}
Here, $\eps$ is the typical energy scale, while $2^{1/6}\sigma$ represents the value where the force vanishes. Using this definition, the force is purely repulsive and vanishes continuously at $ n = n_c = 2^{1/6}\sigma$. Consequently, we define the interaction range of the force as $ n_c$, such that configurations with $n \lesssim \sigma$ are strongly penalized. In this way, we introduce an effective particle diameter.

Chiral activity introduces an additional wall force $\FF^{\rm{w}}_{\rm{a}}$ as a consequence of the particle rotations, which continuously injects angular momentum from the wall to the particle. 
Specifically, chiral active colloids rotating in a fluid or chiral active granular spinners subject to tangential friction forces are subject to additional effective forces that act transversely compared to the distance between the particle and the wall. This force is non-conservative and can be expressed as the cross product between the unit vector normal to the plane of motion, $\ez$, and the direction normal to the wall profile, $\en(s)$:
\begin{equation}
    \FF^{\rm{w}}_{\rm{a}}(n) = -(\ez \times \en(s)\,) f_{\rm{a}}(n)\,\Theta(2n_{c}-n)\,\rm{.}
\label{eq:wall_force_s}
\end{equation}
Here, $f_{\rm{a}}(n)$ denotes the functional form of this force, for which we have chosen a linear profile
\beq
f_{\rm{a}}(n)=\alpha_{\rm{o}}\left(2n_{c} -n\right)\,,
\label{eq:transverse_force_shape}
\eeq
that depends on the distance between the particle and the wall.
The force~\eqref{eq:wall_force_s} is linear and is truncated in the bulk profile as shown in Fig.~\ref{fig:Fig0}(b) and Fig.~\ref{fig:Fig0}(f) for planar geometry and a circular confinement, respectively. 
The cross product implies that $\FF^{\rm{w}}_{\rm{a}}(n)$ is directed tangentially to the wall profile, with a given sign determined by that of the constant  $\alpha_{\rm{o}}$. In addition, this constant determines the strength of this additional non-conservative wall force and can be directly linked to the particle chirality. 
As stated above, $f_{\rm{a}}$ effectively arises from a coarse-graining of microscopic mechanisms, such as hydrodynamic interactions or tangential friction, and is therefore expected to decrease with the wall distance $n$. For simplicity, we adopt a linear dependence, as our results remain qualitatively unchanged for other monotonic functional forms.
In the following sections, we show that the interplay between inertia, curvature, and odd interactions can give rise to density accumulation near the walls, as well as persistent edge currents.

\section{Density accumulation induced by wall curvature}\label{ch:WallAccumulation}

\begin{figure*}[t]
\includegraphics[width=0.96\textwidth, trim=0 0 0 0, clip=true]{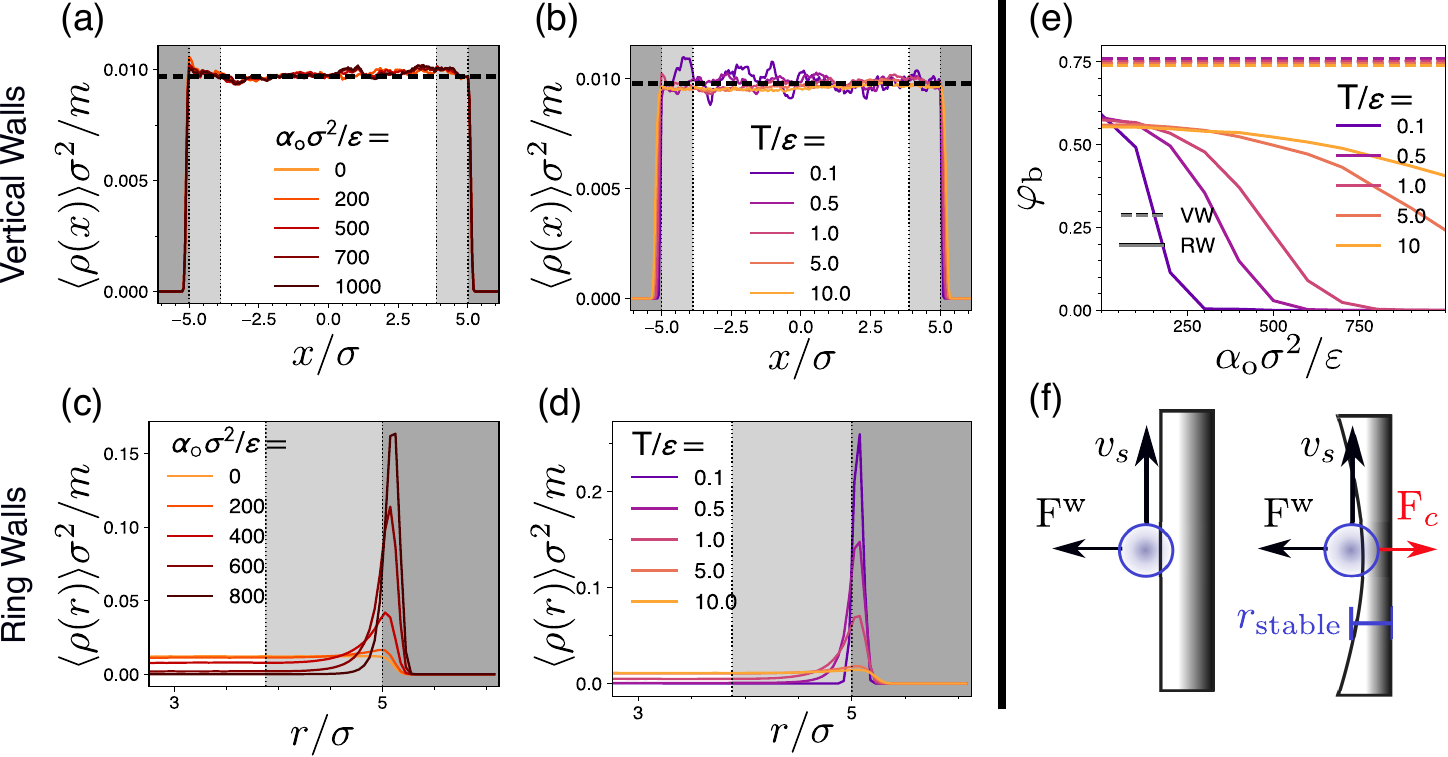}
\caption{ {\textbf{Wall Accumulation in Curved Geometries}}. Figures~(a)-(b) show the average density field $\langle\rho\rangle$ along the $\hat{x}$ direction, corresponding to the wall-normal direction in the vertical walls geometry. The dotted lines represent the average bulk value. Figures (c)-(d) show the same average field for the ring walls geometry. Figures~(a) and (c) display the dependence of the average density on the chirality strength $\alpha_{\rm{o}}\sigma^{2}/\eps$, while Figs.~(b) and (d) show its dependence on the temperature $T/\eps$. Figure~(e) summarizes the behaviour of the average fraction of particles in the bulk, $\varphi_{b}$, for both geometries: vertical walls (dotted lines) and ring walls (solid lines). The bulk fraction is plotted as a function of the chirality strength for different temperatures. 
Figure~(f) schematically illustrates the stabilization mechanism for chiral active particles: due to the tangential velocity $v_{s}$ with respect to the wall, particles moving along a curved profile (right) acquire a centrifugal force $\mathrm{F}_{c}$, which can balance the repulsive wall interaction $\mathrm{F}^{\rm{w}}$ and stabilize the particles at an orbiting distance $r_{\rm{stable}}$. Such a balancing force is absent in a flat wall (left). 
}
\label{fig:Fig1}
\end{figure*}

\begin{figure*}[t]
\includegraphics[width=0.96\textwidth, trim=0 0 0 0, clip=true]{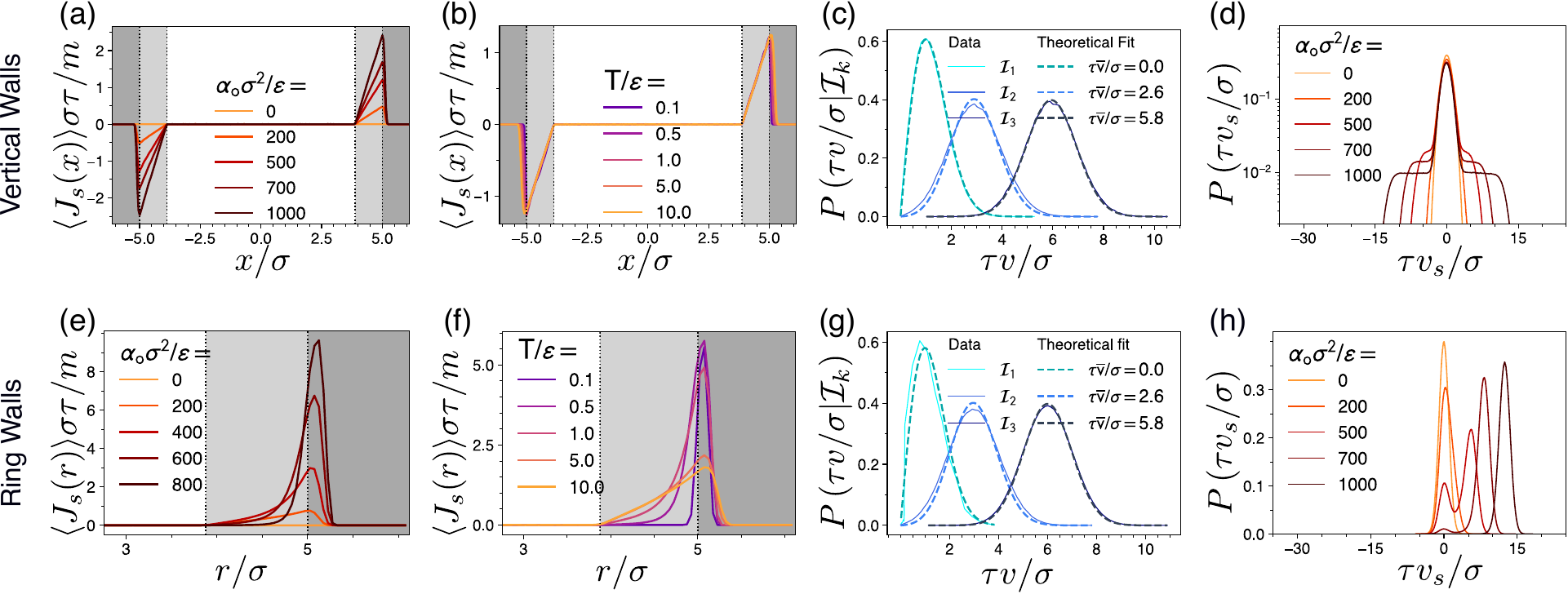}
\caption{ {\textbf{Edge Currents}}. Figures~(a)-(b) show the average transverse mass current $\langle J_{s}(x)\rangle$ along the $\hat{x}$ direction, corresponding to the wall-normal direction in the vertical walls (vertical wall) geometry. Figures~(e)-(f) show the same observable ($\langle J_{s}(r)\rangle$) for the ring walls (ring wall) geometry. Figures~(a) and (e) display the dependence of the current profile on the chirality strength $\alpha_{\rm{o}}\sigma^{2}/\eps$, while Figs.~(b) and (f) show its dependence on the temperature $T/\eps$.
Figures~(c) and (g) display the conditional speed distributions at different distances from the wall, denoted by $\mathcal{I}_{k}$. We display with solid lines the empirical distributions, while dotted lines represent the fit of the empirical curves using $p(v | \overline{v}, T) \propto v \exp{\left(-m(v-\overline{ v})^{2}/(2T)\right)}$, where $T$ is chosen as the bath temperature appearing in the Langevin equation Eq.~\eqref{eq:langevin_eq} and the fit parameters are the normalization constant and $\overline{v}$, whose value is reported in the legend. Here, $\mathcal{I}_{1}$ corresponds to the bulk region, where wall forces are absent; $\mathcal{I}_{2}$ to the region where only chiral forces are present; and $\mathcal{I}_{3}$ to the region where both chiral and repulsive forces act. The similarity between the theoretical fits of Figs.~(c) and (g) suggests that the local dynamics are unaffected by curvature.
Figures~(d) and (h) show the global distributions of the transverse velocity $v_{s}$ for the vertical wall and ring wall geometries, respectively, at different chirality strengths. In the vertical wall geometry, increasing chirality produces broader tails in the distribution, reflecting the presence of chiral interactions localized near the walls. In contrast, the ring wall geometry develops an additional peak, signaling the emergence of wall accumulation through a population of particles with stable transverse velocities. Physical quantities are rescaled using the mass $m$ of the particle, the particle's effective diameter $\sigma$ and the time scale $\tau= \sigma\sqrt{m/\eps}$.
}
\label{fig:Fig2}
\end{figure*}

We implement numerically the dynamics described in Sec.~\ref{ch:Model}, starting from a homogeneous initial configuration in both the vertical-wall and ring-wall geometries (Figs.~\ref{fig:Fig0}(c) and (g)). Specifically, we simulate a population of $N$ independent, non-interacting particles while varying the reduced temperature $T/\eps$ and the amplitude of the chiral active force, $\alpha_{\rm{o}}\sigma^{2}/\eps$. We use $\sigma$, $\varepsilon$, and $m$ as units of length, energy, and mass, respectively, which define the time unit $\tau = \sigma\sqrt{m/\varepsilon}$.

For the vertical wall geometry (Fig.~\ref{fig:Fig0}(d)\,), particles are homogeneously distributed, as in a passive system. 
To quantitatively confirm spatial homogeneity, we compute the average density field $\langle\rho(x)\rangle$ along the wall-normal direction, as shown in Figs.~\ref{fig:Fig1}(a)-(b) for different values of the chirality strength $\alpha_{\rm{o}}\sigma^{2}/\eps$ and reduced temperature $T/\eps$. The density $\langle\rho(x)\rangle$ is constant in the bulk region of the system, while it decreases in the wall region due to repulsive effects.
This spatial distribution remains unchanged when varying $\alpha_{\rm{o}}\sigma^{2}/\eps$ (Fig.~\ref{fig:Fig1}(a)\,) and $T/\eps$ (Fig.~\ref{fig:Fig1}(b)\,).

On the other hand, in a curved geometry, chiral particles tend to accumulate in the wall region. 
This effect is quantitatively shown in Figs.~\ref{fig:Fig1}(c)-(d) by plotting the average density profile $\langle\rho(r)\rangle$ along the wall-normal direction, corresponding to the radial direction pointing away from the center of the circular box.
In this case, the density $\langle\rho(r)\rangle$ displays a peak in the proximity of the wall (dark-grey region) and features an almost constant profile in the bulk (white region). This wall-accumulation phenomenon is driven by chirality and disappears in the non-chiral limit, where the peak is suppressed and the density field becomes uniform, as expected for a passive particle. 
The height of the peak and the strength of this wall-accumulation phenomenon increases with chirality (Fig.~\ref{fig:Fig1}(c)\,) and decreases with temperature  (Fig.~\ref{fig:Fig1}(d)\,).

The accumulation phenomenon observed in our case differs from the wall accumulation obtained for active, self-propelled particles moving at constant speed.
Indeed, in the case of chiral particles interacting with walls through transverse forces, this wall accumulation is induced by wall curvature. 

A heuristic interpretation of this phenomenon can be obtained by considering the force balance in Eqs.~\eqref{eq:wall_force_n}-\eqref{eq:wall_force_s} and neglecting thermal noise. When a particle enters the wall region (dark-grey region in Figs.~\ref{fig:Fig0}(a)-(e)\,), it experiences a repulsive force $\mathrm{F}^{\rm{w}}$ normal to the wall profile, which pushes the particle into the bulk. A chiral particle is also subject to a transverse force responsible for a velocity $v_{s}$ tangential to the wall profile, which induces a net transverse current. This current, later referred to as edge current, is illustrated in Fig.~\ref{fig:Fig1}(f) for both geometries.

The geometry of the wall plays a different role at this stage: if the wall has no curvature (left schematic of Fig.~\ref{fig:Fig1}(f)\,), the particle position is determined by the competition between thermal noise and repulsive wall forces $\mathrm{F}^{\rm{w}}$, since the tangential motion is decoupled from the normal one.
In a curved geometry, this is no longer true. Indeed, the change in direction of the tangential motion gives rise to a centrifugal force ($\mathrm{F}_{c}$) pointing towards the wall. This contribution balances the repulsive force close to the wall profile, creating preferential positions for the particles ($r_{\rm{stable}}$, identified as the position of the density peak) and thus leading to wall accumulation.

In this accumulation process, temperature plays the opposite role to chirality, decreasing the height of the peak, as shown in Fig.~\ref{fig:Fig1}(d).
Indeed, thermal agitation allows particles to escape from the preferential position ($r_{\rm{stable}}$) to the bulk region, thereby weakening the accumulation mechanism. 

We further quantify this behavior by tracking the average fraction of particles in the bulk region, $\varphi_{b}$, as a function of chirality $\alpha_{\rm{o}}$ and temperature $T$. In Fig.~\ref{fig:Fig1}(e), solid curves correspond to the ring-wall geometry and dashed curves to the vertical-wall geometry. Temperature plays qualitatively different roles in the two cases. In the vertical-wall geometry, temperature leaves $\varphi_{b}$ almost unchanged, as expected due to the lack of wall accumulation. In contrast, in the ring-wall geometry, increasing temperature reduces wall accumulation and allows particles accumulated near the wall to re-enter the bulk, thereby increasing $\varphi_{b}$. We remark that, with increasing chirality or decreasing temperature, wall accumulation progressively increases (as evidenced by the decrease of $\varphi_{b}$), until it becomes so strong that it completely depletes the bulk ($\varphi_{b}\to 0$,  scaling with the inverse of the system size). 


\section{Chirality-induced edge currents}\label{ch:EdgeCurrents}

In our system, we expect edge currents due to the presence of the chiral wall forces (Eq.~\eqref{eq:wall_force_s}\,) both for vertical-wall (Fig.~\ref{fig:Fig2}(a)\,) and ring-wall (Fig.~\ref{fig:Fig2}(e)\,) geometries. Since the expression of the active force is non-vanishing in a stripe of width $2n_{c}$, we expect edge currents to be localized near the wall region.

To quantify the edge currents, we compute the average mass current tangential to the wall profile. This average current is calculated at a given distance $n$ from the wall profile, $\langle J_{s}(n)\rangle = \langle \rho(n)v_{s}(n)\rangle$, where $\rho(n)$ is the microscopic mass density field while $v_{s}(n)$ is the microscopic velocity field along the tangent direction to the wall profile. Such a definition holds for both the vertical-wall and ring-wall geometries and can be computed numerically by binning the system.
In both cases, we observe a localized  current in the wall region since $\langle J_{s}\rangle$ vanishes inside the bulk region and displays a peak near the wall. Repulsive effects from the wall then reduce the number of particles close to the boundary and, as a consequence, the mass current decreases again after reaching the peak.
The height of the current peak increases with chirality, as shown in Figs.~\ref{fig:Fig2}(a)-(e). We can already appreciate the difference between the two geometries by examining how $\langle J_{s}\rangle$ approaches its maximum value in Figs.~\ref{fig:Fig2}(a)-(e). Since the chiral force $f_{\rm{a}}$ depends linearly on the wall-distance (Eq.~\eqref{eq:transverse_force_shape}), we expect the mass current to increase in the same way before reaching the peak. This is indeed what occurs in the absence of wall accumulation, namely for vertical walls. In curved geometries, such as ring walls, curvature induces density accumulation, as discussed in Sec.~\ref{ch:WallAccumulation}, thereby modifying the profile of $\langle J_{s}\rangle$, as displayed in Fig.~\ref{fig:Fig2}(e).

Regarding the role of temperature, we observe that it has only a weak effect on the edge currents in the flat geometry (Fig.~\ref{fig:Fig2}(b)\,), whereas it significantly alters the transverse density-current profile in curved geometries ((Fig.~\ref{fig:Fig2}(f)\,). In particular, increasing the temperature lowers the peak by mitigating the accumulation effect and facilitating the re-entry of particles into the bulk, as predicted by our heuristic interpretation for wall accumulation (Sec.~\ref{ch:WallAccumulation}\,).

In addition to edge currents, chirality affects the probability distribution of the speed $v$, which we have monitored conditioned on three spatial regions: inside the bulk, $\mathcal{I}_{1}$, where no wall force is present; in the region where only chiral forces are present, $\mathcal{I}_{2}$; and in the third region, $\mathcal{I}_{3}$, where both $\mathrm{F}^{\rm{w}}_{\rm{a}}$ and $\mathrm{F}^{\rm{w}}$ are present (these three regions correspond respectively to the white, the light-grey and the dark-grey shaded areas of Figs.~\ref{fig:Fig2}(a),(e)\,). Figures~\ref{fig:Fig2}(c),(g) show the conditional distributions for both geometries for a given value of chirality $\alpha_{\rm{o}}$ and temperature $T$.
We fit the numerical speed distributions (solid lines) in the $\mathcal{I}_{k}$ regions using a Gaussian distribution with non-zero mean for the velocity modulus $v = |\vv|$, i.e., $p(v \mid \overline{v}, T) \propto v \exp{\left(-m(v-\overline{v})^{2}/(2T)\right)}$. As a first approximation, we take $T$ to be the bath temperature appearing in the Langevin equation (Eq.~\eqref{eq:langevin_eq}\,). Therefore, the only fitting parameters are the velocity $\overline{v}$ and the normalization constant. The results of the fit are shown as dotted lines in Figures~\ref{fig:Fig2}(c),(g).
The two geometries display the same qualitative behavior: in the bulk, the speed follows the Maxwell–Boltzmann distribution with $\overline{v}=0$, while in the wall regions the mean becomes non-zero and increases as the position approaches the wall. Moreover, by comparing the theoretical fits for the two geometries (dashed lines in Figs.~\ref{fig:Fig2}(c),(g)\,), we find that the fitted values of $\overline{v}$ are consistent across the two geometries. Hence, both vertical walls and ring walls are described by the same distribution $p(v \mid \overline{v}, T)$. For instance, in the $\mathcal{I}_{2}$ region both geometries display a value of $\tau \overline{v} / \sigma \approx 2.6$, while in the $\mathcal{I}_{3}$ region $\tau \overline{v} / \sigma \approx 5.8$. Since $\overline{v}$ is the only fitting parameter apart from the normalization, we can conclude that the distributions are the same in the two geometries.

Despite the different geometries, Figs.~\ref{fig:Fig2}(c) and (g) suggest that the conditional probability distributions can be treated similarly at the local level, thus the curvature plays a marginal role in the local dynamics. However, the density field differs significantly between the two geometries because curvature and chirality induce wall accumulation, which in turn modifies the global velocity statistics.
Figures~\ref{fig:Fig2}(d) and (h) show the total distribution of the transverse velocity for both geometries at different chirality strengths. These distributions reveal the role of chiral interactions even if spatial information is integrated out. In the vertical-wall geometry (Fig.~\ref{fig:Fig2}(d)\,), the distribution is Gaussian in the absence of chirality, while increasing $\alpha_{\rm{o}}$ generates progressively broader tails, signaling a non-equilibrium mechanism localized near the walls. 

In contrast, the ring-wall geometry (Fig.~\ref{fig:Fig2}(h)\,), starting from a Gaussian distribution, develops an additional peak as chirality increases. This peak is centered at the maximum value of the tangential velocity. This qualitative change indicates that the out-of-equilibrium interactions can no longer be disregarded, even though they originate as surface effects. Such behaviour provides further evidence of wall accumulation in curved-wall geometries with chirality, since the population of accumulated particles can become comparable to that of the bulk.

\section{Hydrodynamic theory for edge currents and curvature-induced wall accumulation}\label{ch:Theory}

We now show that the model described by Eq.~\eqref{eq:langevin_eq} (Sec.~\ref{ch:Model}) exhibits density accumulation when curved wall profiles are considered. We derive this phenomenon by constructing a hydrodynamic theory associated with Eq.~\eqref{eq:langevin_eq} for a generic curved wall profile, $\ww$.

To develop a unified theoretical description for both the vertical-wall and ring-wall geometries, we parametrize the wall profile by its arc length, $\ww(s)=(x(s),y(s))$, where $s$ is the arc-length coordinate. At each point of the profile, we define the tangent and normal unit vectors, $\et(s)$ and $\en(s)$, respectively (i.e., $\et=\ey$ and $\en=-\ex$ for the right wall in the vertical-wall geometry, $\et=-\ey$ and $\en=\ex$ for the right one, while $\et=\etheta$ and $\en=-\er$ for the ring-wall geometry). This coordinate system is known as the tubular, or Frenet-Serret, reference frame (see App.~\ref{app:TubularCoordinates} for a detailed description of this coordinate system). It naturally introduces the local curvature $\kappa(s)$, which vanishes for vertical walls and is constant, $\kappa=1/R$, for a ring of radius $R$.
We can build the hydrodynamics starting from the Fokker-Planck equation associated with Eq.~\eqref{eq:langevin_eq} and projecting onto the velocity moments, as shown in App.~\ref{app:Hydrodynamics}. 
The stationary solutions of the hydrodynamic equations can thus be expressed in tubular coordinates, where $n$ is the wall-normal coordinate and $s$ is the arc-length parameter:
\begin{align}
&\partial_{s}\left(\rho u_{s}\right) +\partial_{n}\left((1-n\kappa(s))\rho u_{n}\right) = 0
\label{eq:continuity_eq}\\
&-\rho\dfrac{\kappa(s)}{(1-n\kappa(s))} u_{s}u_{n}
=
\dfrac{\rho}{m} \,\mathrm{F}^{\rm{w}}_{\rm{a}}
-\dfrac{\gamma}{m}\rho u_{s}
\label{eq:momentum_eq_s}\\
&\rho\dfrac{\kappa(s)}{(1-n\kappa(s))}u_{s}^{2}=-\dfrac{T}{m}\partial_{n}\rho
+\dfrac{\rho}{m} \,\mathrm{F}^{\rm{w}}
-\dfrac{\gamma}{m}\rho u_{n}
\label{eq:momentum_eq_n}\,\rm{,}
\end{align}
where we have neglected the advection gradient terms in the momentum equations, since we are interested in the slowly varying hydrodynamic fields. Moreover, we assume the ideal-gas equation of state, $P = T \rho(n)/m$ since our description will focus on non-interacting particles. 
The left-hand sides of Eqs.~\eqref{eq:momentum_eq_s},\eqref{eq:momentum_eq_n} represent transport terms arising from curvature effects and contain two additional geometric contributions that are not explicit in Cartesian coordinates and become evident from the use of the curvilinear reference frame.  In particular, Eq.~\eqref{eq:momentum_eq_s} contains a Coriolis-like term proportional to $u_s u_n$, while Eq.~\eqref{eq:momentum_eq_n} includes a centrifugal-like contribution proportional to $u_s^2$. These terms are induced by the local rotation of the frame (following the wall profile) and vanish identically for a straight wall ($\kappa=0$). These terms highlight the role of curvature and transverse motion, as they couple the density field to the tangential velocity.

Such a representation helps disentangle the wall-particle interaction, separating the repulsive term acting along the wall-normal direction from the chiral active force, $\mathrm{F}^{\rm{w}}_{\rm{a}}$, which acts along the tangential direction of the wall profile (we recall that $-\ez \times \en(s) = \et(s)$\,).

In order to model the impermeability condition within the walls, we can put $u_{n}=0$ in Eqs.~\eqref{eq:momentum_eq_s}-\eqref{eq:momentum_eq_n}. Under this approximation Eq.~\eqref{eq:momentum_eq_s} decouples completely from Eq.~\eqref{eq:momentum_eq_n}, yielding a closed relation for $u_{s}$:
\beq
u_{s}=\dfrac{1}{\gamma}\,\mathrm{F}_{\rm{a}}^{\rm{w}}\,\rm{,}
\label{eq:edge_currents_tc}
\eeq
which can be used in order to determine the intensity of the edge currents, numerically observed in Sec.~\ref{ch:EdgeCurrents}. Substituting this expression into Eq.~\eqref{eq:momentum_eq_n} yields a closed equation for the mass density field $\rho$:
\beq
\begin{split}
\dfrac{T}{m\gamma}\partial_{n}\ln(\rho)
= &
-\dfrac{1}{\gamma m}\partial_{n}U^{\rm{w}}(n)\Theta(n_{c}-n) \\
&-\dfrac{\kappa(s)\alpha_{\rm{o}}^{2}}{(1-n\kappa(s))\gamma^{3}}
(n-2n_{c})^{2}\Theta(2n_{c}-n)\,\rm{.}
\end{split}
\label{eq:closed_un_tc}
\eeq
Here we have used Eqs.~\eqref{eq:wall_force_n}-\eqref{eq:wall_force_s}. Notice that the direction of the edge current, determined by the sign of $\alpha_{\rm{o}}$, does not affect $\rho$ because it enters Eq.~\eqref{eq:closed_un_tc} quadratically. Moreover, the dependence of $\rho$ on the arc-length coordinate $s$ arises exclusively through the curvature.

\begin{figure}[t]
\includegraphics[width=0.98\columnwidth, trim=0 0 0 0, clip=true]{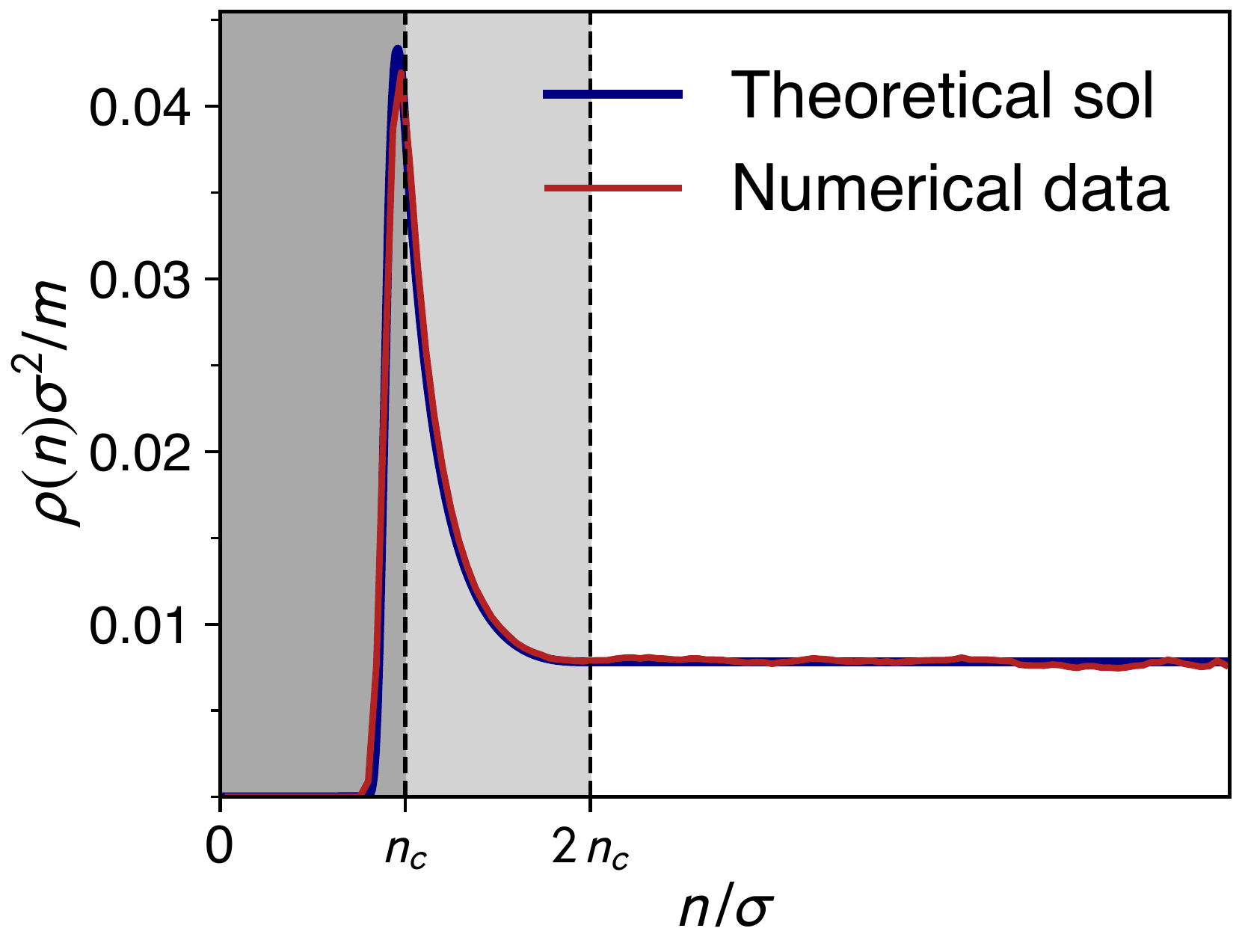}
\caption{{\textbf{Theoretical profile of the average density}}. The graph shows the solution $\rho(n)$ (blue curve) of Eq.~\eqref{eq:closed_un_tc} as a function of the wall-particle distance along the normal direction $n$ in the three different regions, compared with the numerical result (red curve). The first region (dark-grey shaded) is where tangential ($\mathrm{F}_{\rm{a}}^{\rm{w}}$) and repulsive forces ($\mathrm{F}^{\rm{w}}$) are non-vanishing. Then, ofr $n_{c}\leq n \leq 2n_{c}$ (light-grey shaded), only chiral forces are present, and finally inside the bulk (white) region, no wall forces are present. The curve solves the  Eq.~\eqref{eq:closed_un_tc} with boundary condition $\rho(n_{b},t)=\rho_{b}$. The curve is obtained with $\rho_{b}\sigma^2/m \approx 0.01$, $\alpha_{\rm{o}}\sigma^{2}/\eps = 500$, $T/\eps = 1$, $R \simeq 6.12\,\sigma$, $\tau\gamma/m = 100$ (where $\tau = \sigma\sqrt{m/\varepsilon}$ is the characteristic time of the system), $\kappa = R^{-1}$, and $n_{c}=2^{1/6}\sigma$.}
\label{fig:Fig1th}
\end{figure}

Equation~\eqref{eq:closed_un_tc} then becomes a first-order differential equation for the density profile, to be solved between the wall ($n=0$) and the bulk ($n=n_{b}>2n_{c}$), with boundary condition $\rho(n_{b},s)=\rho_{b}(s)$. 
The solution reads:
\begin{equation}
\begin{split}
&\rho(n,s) = \rho_{b}(s)\,\text{exp}\Bigg( -\dfrac{1}{T} U^{\rm{w}}(n)\Theta(n_{c}-n) \\
&+ \dfrac{m\alpha_{\rm{o}}^{2}}{T\gamma^{2}}  \Theta(2n_{c}-n) \Bigg[ \dfrac{(2n_{c}-n)^{2}}{2} - \dfrac{1-2n_{c}\kappa(s)}{\kappa(s)}(2n_{c}-n) \\
&+ \left(\dfrac{1-2n_{c}\kappa(s)}{\kappa(s)}\right)^{2} \ln{\left(1 + \dfrac{\kappa(s)(2n_{c}-n)}{1-2n_{c}\kappa(s)}\right)} \Bigg]\Bigg) \,\rm{.}
\label{eq:rho_solution}
\end{split}
\end{equation}
Equation~\eqref{eq:rho_solution} shows (blue curve in Fig.~\ref{fig:Fig1th}\,) that our theory is able to predict a peak in the density near the wall, together with edge currents (Eq.~\eqref{eq:edge_currents_tc}\,). 

We compare this prediction with numerical results in Fig.~\ref{fig:Fig1th} for the ring-wall geometry, compared with the numerical measurement (red curve in Fig.~\ref{fig:Fig1th}\,). In this case, the rotational symmetry implies that the solution does not depend on the arc-length parameter and is uniquely described by the normal coordinate, $n$. The two curves are compared such that the bulk value $\rho_{b}$ is the same. 

The density profile, $\rho$, remains continuous throughout the wall region. Starting from the bulk and decreasing the radial coordinate $n$ toward the wall, the density first increases in the interval $n_c \leq n \leq 2n_c$, where only the chiral force $\mathbf{F}^{\rm w}_{\rm a}$ is present. This increase leads to an accumulation peak near $n \simeq n_c$. For $0 \leq n \leq n_c$, the repulsive wall force also acts and competes with the chiral accumulation mechanism. Very close to the wall, where $n \ll n_c$, repulsive effects dominate, and the density is strongly suppressed.


The result is in quantitative agreement with the numerical data, even for large chirality ($\alpha_{\rm{o}}$) values. 
Moreover, the result is valid for sufficiently large radii of curvature, corresponding to the regime $2n_{c}\kappa \lesssim 1$. Outside this regime, curvature effects become too strong and the approximations underlying the derivation are no longer expected to hold.

\section{Conclusions}\label{ch:conclusions}
We study chiral active particles in confined geometries, showing that the curvature of the wall profile can induce density accumulation near boundaries. 
This steady-state phenomenon, which is absent in flat geometries, does not require self-propulsion and, thus, differs from the wall accumulation induced by the persistence of motile active particles.
Here, we show that this curvature-induced wall-accumulation phenomenon is sustained by a centrifugal force term that is induced by the edge currents generated by chiral interactions. This effect is weakened by thermal fluctuations, which facilitate the re-entry of particles into the bulk. 
The observed behavior is explained through a hydrodynamic theory for coarse-grained fields, such as density and the momentum field, which highlight the interplay between curvature and chiral forces. This theory yields results in quantitative agreement with numerical simulations.

Our results can be extended to more complex geometries with non-constant curvature where higher-order corrections may become crucial to theoretically explain the results. This effect may represent a novel perspective for designing a new generation of motors, generalizing the bacterial-motor idea proposed a decade ago~\cite{di2010bacterial,sokolov2010swimming,hiratsuka2006microrotary,xu2021rotation,vizsnyiczai2017light}. Extending these results to chiral particles~\cite{li2023chirality} and taking advantage of edge currents may increase the motor efficiency.

The present study focuses on the dilute regime where chiral particles do not interact.
However, it is well-known that the interplay between chirality and interactions gives rise to several collective phenomena~\cite{liebchen2017collective,liao2018clustering,reichhardt2019active,bickmann2022analytical,huang2020dynamical,kruk2020traveling,liao2021emergent,debets2023glassy,kalz2026reversal,kuroda2025singular}, ranging from demixing in
clockwise and counterclockwise mixtures~\cite{scholz2018rotating,levis2019simultaneous}, spontaneous vortices~\cite{shee2024emergent,musacchio2025circling} and hyperuniformity~\cite{zhang2022hyperuniform,kuroda2023microscopic,maire2025hyperuniformity,kuroda2025long,zhou2025visual} in non-aligning systems as well as global synchronization, self-proliferating spiral waves~\cite{uchida2010synchronization,uchida2010synchronizationSecond} and microflock patterns~\cite{levis2018micro} due to interplay between chirality and alignment interactions~\cite{negi2023geometry,kreienkamp2022clustering}.
Chiral particles typically interact with transverse forces~\cite{caporusso2024phase} which have been recognized as a general ingredient responsible for the generation of a bubble phase~\cite{caprini2025bubble}, termed BIO, also observed in the presence of tangential friction~\cite{digregorio2026phase} and hydrodynamic interactions~\cite{shen2023collective}.
Understanding how these collective phenomena affect the curvature-induced wall-accumulation phenomenon discovered here represents a future perspective to address. We remark that the theory can be generalized to interacting systems by incorporating shear and odd viscosity~\cite{fruchart2023odd, markovich2021odd, han2021fluctuating, reichhardt2022active,lou2022odd,mecke2023simultaneous, hosaka2023lorentz}, as well as the torque density~\cite{marini2026emergent,maire2026kinetic} recently discovered to govern chiral active systems. 

The curvature-induced wall accumulation phenomenon calls for experimental verification. Good candidates are granular spinners, generating edge currents as a result of the tangential friction between particles and the wall~\cite{caprini2025active,digregorio2026phase}. Since our results are obtained in the limit of finite but very small inertia, we expect similar phenomena to arise in systems of spinning colloids~\cite{mecke2024chiral,massana2021arrested,han2021programmable}, where edge currents emerge from the hydrodynamic flows generated by particle rotation.

\appendix

\section{Tubular coordinate system}\label{app:TubularCoordinates}
In this Appendix, we introduce the tubular -- also termed Frenet-Serret -- coordinate system, which will be used to derive the main analytical results in App. \ref{app:Hydrodynamics}.
The tubular coordinate system describes a two-dimensional vector through a rotating reference frame along a differentiable curve $\ww(s)=(x(s), y(s)\,)$, where $s$ is the arc-length parameter along the curve. This description pictorially resembles a tube whose profile follows the shape of the original curve $\ww(s)$. In the presence of a (curved) wall profile, this reference frame can be convenient compared to the Cartesian one to decouple tangential and normal forces exerted by the wall.

We start by representing the position vector $\rr_{\rm{p}}$ in the Frenet–Serret reference frame. It can be decomposed into two components: the first is tangential to the curve $\ww(s)$, and the second is normal to it:
\beq
\rr_{\rm{p}}(s,n) = \ww(s) + n\,\en (s)\,\rm{.}
\label{eq:tc_embedding_app}
\eeq
Here, $n$ is the coordinate along the normal direction $\en(s)$ of the curve $\ww$.
In these coordinates, the tangential direction to the wall profile can be expressed as $\et(s) = \partial_{s}\ww(s)$ that, together with $\en(s)$, forms an orthogonal basis. Using that $\et(s) \cdot \et(s) = 1 = \en(s) \cdot \en(s)$ and $\et(s) \cdot \en(s) = 0$, one can obtain, by differentiated along $s$, the Frenet-Serret formulas \cite{Gray2004Tubes}:
\begin{align}
\partial_{s} \et(s) &= \kappa(s) \en(s) \label{eq:partial_s_es} \\
\partial_{s} \en(s) &= -\kappa(s) \et(s) \label{eq:partial_s_en}\,\rm{,}
\end{align}
where $\kappa(s) = |\partial_{s}\et(s)|$ is defined as the curvature of $\ww$ and it is the inverse of the radius of curvature $R_{\kappa}(s)=\kappa(s)^{-1}$.

By differentiating $\rr_{p}$ with respect to time and using Eqs.~\eqref{eq:partial_s_es}-\eqref{eq:partial_s_en}, we obtain the expression for the velocity in the Frenet–Serret coordinate system,
\beq
\dot{\rr}_{p} = (1-n\kappa(s))\dot{s}\,\et(s) + \dot{n}\,\en(s)\,\rm{.}
\label{eq:dot_rp}
\eeq
We can also express $\dot{\rr}_{p}$ in terms of its tangential and normal components with respect to the wall profile, namely $\dot{\rr}_{p} = v_{s} \et(s) + v_{n}\en(s)$.
Comparing this expression with the right-hand side of Eq.~\eqref{eq:dot_rp}, we identify the following equations for the spatial coordinates:
\begin{align}
\dfrac{d s}{dt} =&\, \dfrac{1}{1-n\kappa(s)}v_{s}
\label{eq:dot_s}\\
\dfrac{d n}{dt} =&\, v_{n}
\label{eq:dot_n}\,\rm{.}
\end{align}
In order to derive the dynamics of the tangential and normal component velocity, $v_s$ and $v_n$, we compute $\ddot{\rr}_{p}$, by differentiating Eq.~\eqref{eq:dot_rp} with respect to time and use Eqs.~\eqref{eq:partial_s_es}-\eqref{eq:partial_s_en} to calculate $\partial_{s} \et(s)$ and $\partial_{s} \en(s)$.
Then, we need to eliminate $\dot{n}$ and $\dot{s}$ through Eqs.~\eqref{eq:dot_s} and ~\eqref{eq:dot_n}, as well as $\ddot{s}$ by using the time derivative of the expression~\eqref{eq:dot_s}, which reads
\begin{equation}
    \ddot{s} = \frac{1}{1-n \kappa(s)}\dfrac{d v_{s}}{dt} + \frac{v_s}{(1-n \kappa(s))^2} \left( \dot{s} n\partial_{s} \kappa(s)  + \kappa(s) \dot{n}\right) \,.
\end{equation}
By projecting the force terms in the right-hand-side of Eq.~\eqref{eq:langevin_eq} on the tangential ($\propto \et(s)$) and wall-normal ($\propto \en(s)$) directions, we obtain
%
\begin{align}
m\dfrac{d v_{s}}{dt} = &-\gamma v_{s} + \sqrt{2\gamma T}\,\xi_{s}(s,t) + \mathrm{F}^{\rm{w}}_{\rm{a}}(n) \nonumber \\
& + m\dfrac{\kappa(s)}{1-n\kappa(s)}v_{s}v_{n}
\label{eq:dot_vs} \\
m\dfrac{d v_{n}}{dt} = &-\gamma v_{n} + \sqrt{2 \gamma T}\,\xi_{n}(s,t) + \mathrm{F}^{\rm{w}}(n) \nonumber \\
&- m\dfrac{\kappa(s)}{1-n\kappa(s)}v_{s}^{2}\,\rm{,}
\label{eq:dot_vn}
\end{align}
where the terms $\xi_{s}$ and $\xi_{n}$ in Eqs.~\eqref{eq:dot_vs}-~\eqref{eq:dot_vn} are, respectively, the projections of a Gaussian white noise onto the tangential and wall-normal directions and therefore depend on the arc-length parameter $s$. The tubular representation explicitly reveals new terms in the dynamics that arise due to curvature. In particular, the last term in Eq.~\eqref{eq:dot_vs} accounts for the Coriolis-like acceleration, while the last term on the right-hand side of Eq.~\eqref{eq:dot_vn} corresponds to the centrifugal-like contribution.
The advantage of choosing such a reference frame lies in the fact that the wall force, $\FF^{\rm{w}}$, and the chiral active interaction, $\FF^{\rm{w}}_{\rm{a}}$, become separated in the dynamics, since the former acts along the normal direction to the wall, whereas the latter acts along the tangential direction ( $-\ez \times \en(s) = \et(s)$\,). Moreover, we recall that both terms depend solely on their distance to the wall $n = |\ww- \rr|$, according to Eqs.~\eqref{eq:wall_force_n},~\eqref{eq:wall_force_s}.

On the other hand, we can differentiate $\rr_{p}$ with respect to the tubular variables and use Eqs.~\eqref{eq:partial_s_es}-\eqref{eq:partial_s_en} in order to obtain the expression for the metric tensor $g_{ij}$:
\beq
g_{ij} = \partial_{i}\rr_{\rm{p}} \cdot \partial_{j}\rr_{\rm{p}} = \begin{pmatrix}
(1-n\kappa(s))^{2} & 0 \\
0 & 1
\end{pmatrix} \,\rm{,}
\label{eq:metric_tc}
\eeq
where the indices span the tangential and normal coordinates, $i,j=\{s,n\}$. From here on, we use Latin indices to denote the coordinates $s$ and $n$.
The line
element is given by $d \ell^{2} = (1-n\kappa(s))^{2}ds^{2} + dn^{2}$ while the Jacobian has the simple form $\sqrt{g} = \sqrt{\det{g}} = (1-n\kappa(s))$. As a consequence, the phase-space measure in the new coordinate system becomes $(1-n\kappa(s)\,)ds\,dn\,dv_{s}\,dv_{n}$.

Since the metric is diagonal, the inverse matrix $g^{ij}$ can be obtained simply by taking the reciprocal of the diagonal elements. Moreover, it is useful to define the stretching factors along each direction, namely $h_{i}=\sqrt{g_{ii}}$:
\beq
h_{s} = (1-n\kappa(s))\,, \quad h_{n} = 1\,\rm{.}
\eeq
These factors set the scaling between the coordinate of a vector and its physical (geometric) component. For instance for the  velocity, the stretching factors are the proportionality constants between $\dot{s}$ and $v_{s}$, and $\dot{n}$ and $v_{n}$. These stretching factors account for the geometric deformation induced by the change of variables.
We can now compute the Christoffel symbols using
\beq
\Gamma^{k}_{ij} = \dfrac{1}{2}g^{kl}(\partial_{i}g_{jl} + \partial_{j}g_{il} - \partial_{l}g_{ij}) \, \rm{,}
\label{eq:Christoffel_symbols_def}
\eeq
where we have omitted the summation over repeated indices using the Einstein convention. Here, $g^{kl}$ is the inverse of the metric tensor $g$ defined in Eq.~\eqref{eq:metric_tc}, i.e. $g^{kl} = (g^{-1})_{kl}$. Using Eq.~\eqref{eq:Christoffel_symbols_def} together with Eq.~\eqref{eq:metric_tc}, we find that only four out of eight elements of $\Gamma^{k}_{ij}$ are non-vanishing:
\begin{align}
\Gamma^{s}_{ss} &= -\dfrac{n\partial_{s}\kappa(s)}{1-n\kappa(s)} \\
\Gamma^{s}_{sn} &= -\dfrac{\kappa(s)}{1-n\kappa(s)} = \Gamma^{s}_{ns} \\
\Gamma^{n}_{ss} &= \kappa(s)(1-n\kappa(s))\,\rm{.}
\label{eq:Christoffel_symbols_tc}
\end{align}
This framework is useful for expressing differential operators in the tubular coordinate system starting from their Cartesian counterparts. For instance, for a generic scalar function $p$ and vector field $\uu$, we have
\begin{align}
(\bfnabla p)^{i} &= \partial^i p = g^{ij}\partial_{j} p \label{eq:covariant_scalar_gradient}\\
\nabla_{i} u^{j} & = \partial_{i}u^{j}+ \Gamma^{j}_{ik}u^{k}\label{eq:covariant_derivative}\\
\bfnabla \cdot \uu & = \dfrac{1}{\sqrt{g}}\partial_{i}(\sqrt{g}u^{i}) \label{eq:covariant_divergence}\,\rm{,}
\end{align}
which represent, respectively, the covariant gradient of a scalar field $p$, as well as the covariant derivative and divergence of a vector field $\uu$. We will make use of these expressions in App.~\eqref{app:Hydrodynamics}.

\section{The Hydrodynamics of non-interacting chiral active particles in tubular coordinates
}\label{app:Hydrodynamics}
In this Appendix, we derive Eqs.~\eqref{eq:continuity_eq}–\eqref{eq:momentum_eq_n} of Sec.~\eqref{ch:Theory}, which provide the hydrodynamic description of our dynamics in tubular coordinates. We first obtain the hydrodynamic equations in Cartesian coordinates and then transform them to the Frenet–Serret frame (see App.~\ref{app:TubularCoordinates} for further details).

We start from the underdamped Langevin equation for the velocity field $\vv= \dot{\rr}$, describing our chiral non-motile particle (Eq.~\eqref{eq:langevin_eq}\,). We can write the associated Fokker-Planck equation for the single-particle probability density $f= f(x,y,v_{x}, v_{y},t)$:
\begin{flalign}
\partial_{t}f &+ \bfnabla_{\rr}\cdot \left(\vv f\right) + \bfnabla_{\vv}\cdot \left[ \left(-\dfrac{\gamma}{m}\vv + \dfrac{\FF^{\rm{w}}(\rr)}{m}+ \dfrac{\FF^{\rm{w}}_{\rm{a}}(\rr)}{m}\, \right)f\right] \nonumber\\
&= \dfrac{\gamma T}{m^{2}}\bfnabla_{\vv}^{2} f\,\rm{.} 
\label{eq:FPE}
\end{flalign}
A hydrodynamic description can be derived by projecting Eq.~\eqref{eq:FPE} onto the velocity moments of the distribution $f$. Specifically, we define the mass and momentum density fields as
\beq
\begin{pmatrix} \rho(\rr,t) \\  \rho(\rr,t)\uu(\rr,t) \end{pmatrix}\equiv
\int d\vv \, f(\rr,\vv,t) 
\begin{pmatrix} m \\ m\vv  \end{pmatrix} \,\rm{.}
\label{eq:hydrofields}
\eeq
The dynamical equations for these fields can be derived by multiplying both sides of Eq.~\eqref{eq:FPE} by $m$ and $m v^{i}$, respectively.
Upon integrating the Fokker–Planck equation over the velocity variables, we obtain an infinite hierarchy of coupled dynamical equations, known as the moment hierarchy: for instance, the first equation involves the zeroth and first moments, the second equation the first and second moments, and so on.

We assume that higher-order velocity moments do not contribute significantly to the hydrodynamics, as we are interested in slowly varying fields. We therefore close the hierarchy at the level of the second moment, which corresponds to the velocity current defined in Eq.~\eqref{eq:hydrofields}. In particular, we set
$T\rho(\rr,t)/m\, \delta^{ij}= \int d\vv \, m (v^{i}-u^{i})(v^{j}-u^{j})f$, which assumes that, at leading order, the thermal velocity is determined by the bath temperature $T$, which is the same temperature appearing in the Langevin equation~\eqref{eq:langevin_eq}.

Accordingly, the hydrodynamic description of the system reads
\begin{align}
\partial_{t} \rho + \bfnabla \cdot \left(\rho \uu\right) =& 0
\label{eq:continuity_eq_hydro_wall}\\
\rho\left(\partial_{t} \uu + \uu \cdot\bfnabla \uu \right) =& -\bfnabla P(\rr) - \dfrac{\gamma}{m} \rho \uu \nonumber \\
&+ \dfrac{\rho}{m} \left(\FF^{\rm{w}}(\rr) + \FF^{\rm{w}}_{\rm{a}}(\rr)\right) \,\rm{,}
\label{eq:momentum_eq_hydro_wall}
\end{align}
where the pressure tensor is chosen according to the ideal gas law, $P(\rr) = T\,\rho(\rr)/m$. We recall that the velocity field has (contravariant) components $u^{i}$, which are related to the geometrical (physical) components through the stretching factors as $u^{i} = u_{i}/h_{i}$, so that
\begin{align}
    u_n &= u^n \label{eq:physical_un}\\
    u_s &= (1 - n \kappa(s)) u^s
    \,\rm{,} \label{eq:physica_us}
\end{align}
and the same relation holds for other vectorial fields, as the force terms.
From here on, and throughout the main text, we will use upper indices for the components, while lower indices will be used as dummy labels to denote the geometrical components, losing the covariant nature of the theory.

Using the metric tensor, $g_{ij}$ (Eq.~\eqref{eq:metric_tc}), its inverse $g^{ij}$, and the Christoffel symbols, $\Gamma^{k}_{ij}$ (Eq.~\eqref{eq:Christoffel_symbols_tc}), we can express the spatial operators in Eq.~\eqref{eq:continuity_eq_hydro_wall}-\eqref{eq:momentum_eq_hydro_wall} in tubular coordinates (see App.~\eqref{app:TubularCoordinates}\,), obtaining:
\begin{align}
\partial_{t} \rho + \dfrac{1}{\sqrt{g}}\partial_{i}\left(\sqrt{g}\rho u^{i}\right) = &0
\label{eq:continuity_eq_hydro_wall_metric}\\
\rho\left(\partial_{t} u^{i} + u^{j}\partial_{j}u^{i} + \Gamma^{i}_{jk}u^{j}u^{k}\right)
= &-g^{ij}\partial_{j}P -\dfrac{\gamma}{m}\rho u^{i} \nonumber \\
&+ \dfrac{\rho}{m}\left(\mathrm{F}^{\rm{w}}\right)^{i} + \dfrac{\rho}{m}\left(\mathrm{F}^{\rm{w}}_{\rm{a}}\right)^{i}
\label{eq:momentum_eq_hydro_wall_metric}\,\rm{.}
\end{align}
We now explicitly write $\Gamma_{ij}^k$ and $g^{ij}$ in the previous equations and express every contravariant component (upper indices) in terms of the physical components (lower indices), according to Eqs.~\eqref{eq:physical_un} and \eqref{eq:physica_us}. For instance, the continuity equation becomes
\beq
\begin{split}
\partial_{t} \rho &+ \dfrac{1}{1-n\kappa(s)} \Bigg[\partial_{s}\left((1-n\kappa(s))\rho \dfrac{u_{s}}{1-n\kappa(s)}\right) \\
&+ \partial_{n}\left((1-n\kappa(s))\rho u_{n}\right)\Bigg]=0\,\rm{.}
\end{split}
\eeq
For what concerns Eq.~\eqref{eq:momentum_eq_hydro_wall_metric}, we notice that the advection terms $u^{j}\partial_{j}u^{i}$ give rise to additional curvature terms:
\begin{align}
u^{j}\partial_{j}u^{s} &= \dfrac{u_{s}\partial_{s}u_{s}}{(1-n\kappa(s))^{2}} + \dfrac{u_{n}\partial_{n}u_{s}}{(1-n\kappa(s))} + \dfrac{n\partial_{s}\kappa(s)}{(1-n\kappa(s))^{3}} u_{s}^{2} \nonumber \\
&+  \dfrac{\kappa(s)}{(1-n\kappa(s))^{2}} u_{s}u_{n} \\
u^{j}\partial_{j}u^{n} &= \dfrac{u_{s}\partial_{s}u_{n}}{1-n\kappa(s)} + u_{n}\partial_{n}u_{n}\,\rm{.}
\end{align}
We consider stationary solutions of Eqs.~\eqref{eq:continuity_eq_hydro_wall_metric}-\eqref{eq:momentum_eq_hydro_wall_metric} and neglect the velocity-gradient terms $u_{j}\partial_{j}u_{i}$, expressing the equations in terms of the physical components $u_{s}$ and $u_{n}$. By explicitly writing $\Gamma_{ij}^k$ and $g^{ij}$, and after some algebra, the hydrodynamic system becomes:
\begin{align}
&\partial_{s}\left(\rho u_{s}\right) +\partial_{n}\left((1-n\kappa(s))\rho u_{n}\right) = 0
\label{eq:continuity_eq_app}\\
&-\rho\dfrac{\kappa(s)}{(1-n\kappa(s))} u_{s}u_{n}
=
\dfrac{\rho}{m} \,\mathrm{F}^{\rm{w}}_{\rm{a}}
-\dfrac{\gamma}{m}\rho u_{s}
\label{eq:momentum_eq_s_app}\\
&\rho\dfrac{\kappa(s)}{(1-n\kappa(s))}u_{s}^{2}
=
-\dfrac{T}{m}\partial_{n}\rho
+\dfrac{\rho}{m} \,\mathrm{F}^{\rm{w}}
-\dfrac{\gamma}{m}\rho u_{n}
\label{eq:momentum_eq_n_app}\,\rm{,}
\end{align}
which correspond respectively to Eqs.~\eqref{eq:continuity_eq}-\eqref{eq:momentum_eq_n} of Sec.~\ref{ch:Theory}. The advantage of this coordinate system is that it disentangles the normal and tangential motion along a generic wall profile and can therefore be used for both the vertical-wall and ring-wall geometries.



\section*{References}

\bibliographystyle{apsrev4-1}

\bibliography{chiral_walls.bib}

\end{document}